\documentclass[floatfix,showpacs,prl,aps,twocolumn]{revtex4}
\usepackage{bm}
\usepackage{amsmath}
\usepackage{amssymb}
\usepackage{latexsym} 
\usepackage{amsfonts} 
\usepackage{epsfig}
\usepackage{psfrag} 

\newcommand{\ord}{{\mathcal O}} 
\newcommand{\ket}[1]{|#1\rangle} 
\newcommand{\bra}[1]{\langle#1|} 
\newcommand{\bracket}[2]{\langle#1|#2\rangle} 
\newcommand{\nn}{\nonumber\\} 
\newcommand{\bea}{\begin{eqnarray}}
\newcommand{\ea}{\end{eqnarray}} 
 
\newcommand{\pdiff}[2]{\frac{\partial#1}{\partial#2}}

\begin{document}



\title{Decoherence in the dynamical quantum phase transition of the
  transverse Ising chain}  


\author{Sarah Mostame, Gernot Schaller, and Ralf Sch\"utzhold$^*$}

\affiliation{Institut f\"ur Theoretische Physik, 
Technische Universit\"at Dresden, 01062 Dresden, Germany}

\begin{abstract} 
For the prototypical example of the Ising chain in a transverse field,
we study the impact of decoherence on the sweep through a second-order
quantum phase transition.
Apart from the advance in the general understanding of the dynamics of
quantum phase transitions, these findings are relevant for adiabatic
quantum algorithms due to the similarities between them. 
It turns out that (in contrast to first-order transitions studied
previously) the impact of decoherence caused by a weak coupling
to a rather general environment increases with system size 
(i.e., number of spins/qubits), which might limit the scalability of
the system.  
\end{abstract} 


\maketitle
 

Recently, the dynamics of quantum phase transitions \cite{sachdev}  
attracted increasing interest, see, e.g.,
\cite{dziarmaga,cincio+sen,fubini}.  
In contrast to thermal transitions (usually driven by the competition 
between energy and entropy), they are characterized by a fundamental
change of the ground state structure (e.g., from para- to
ferro-magnetic) at the critical value of a variable external parameter 
(e.g., magnetic field).
Quantum phase transitions are induced by quantum rather than thermal
fluctuations and thus may occur at zero temperature.  
At the critical point, the energy levels become arbitrarily close and
thus the response times diverge (in the continuum limit).
Consequently, during the sweep trough such a phase transition by
means of a time-dependent external parameter, small external
perturbations or internal fluctuations become strongly amplified -- 
leading to many interesting effects, 
see, e.g., \cite{group_phase,vidal,zurek,damski,sengupta}.
One of them is the anomalously high susceptibility to decoherence 
(see also \cite{fubini}):
Due to the convergence of the energy levels at the critical point,
even low-energy modes of the environment may cause excitations and
thus perturb the system. 
Here, we study the decoherence caused by a small coupling to
a rather general reservoir for the quantum Ising chain in a transverse
field, which is considered a prototypical example \cite{sachdev}  
for a second-order quantum phase transition 
(and further possesses the advantage of being analytically solvable). 

Apart from the general understanding of quantum phase transitions,
these investigations are also relevant for quantum computing:
By constructing the Hamiltonian appropriately, it is possible to
encode the solution to hard computational problems (such a factoring
large numbers) in its ground state.
In order to reach this solution state, we may start off with a simpler
Hamiltonian whose ground state is easy to prepare as the initial
configuration.
If we now steadily transform it to the problem Hamiltonian, the
adiabatic theorem tells us that we stay near the ground state if the
evolution is slow enough -- and thus finally end up in (or close to)
the desired solution state (adiabatic quantum computing
\cite{farhi,sarandy}).   
However, somewhere on the way from the simple initial configuration to 
the final state, there is typically a critical point which bears
strong similarities to a quantum phase transition 
(e.g., vanishing gap and diverging entanglement in the continuum limit
\cite{latorre,gernot}).
Based on this similarity, it seems \cite{gernot} that adiabatic
quantum algorithms corresponding to second-order quantum phase
transitions should be advantageous compared to isolated avoided level
crossings (which are analogous to first-order transitions).
For an adiabatic quantum algorithm (Grover's search routine \cite{roland}) 
based on a single isolated avoided level crossing \cite{markus}, the
impact of decoherence induced by a low-temperature bath with a
well-behaved spectral distribution does not destroy the scalability of
the system.  
However, as we shall see below, the situation may be very different
for second-order transitions.
These investigations are particularly relevant in view of the recent
announcement (see, e.g., \cite{dwave}) regarding the construction of
an adiabatic quantum computer with 16 qubits in the form of a
two-dimensional Ising model.  


The open system under consideration is described by the total
Hamiltonian ${H}$ which can be split up into that of the closed
system ${H}_{\rm{sys}}$ and the bath ${H}_{\rm{bath}}$ acting
on independent Hilbert spaces  
${\mathfrak H}_{\rm{sys}}\otimes{\mathfrak H}_{\rm{bath}}={\mathfrak H}$ 
\bea
\label{HTotal}
{H}(t) \, = \,  
{H}_{\rm{sys}} (t) + {H}_{\rm{bath}} (t) + 
\lambda \, {H}_{\rm{int}} (t)
\,, 
\ea
plus an interaction $\lambda \, {H}_{\rm{int}}$ between the two,
which is supposed to be weak $\lambda\ll1$ in the sense that it does
not perturb the state of the system drastically.  
Note, however, that the change of the bath caused by the interaction
with the system need not be small. 
In order to describe the evolution of the combined quantum state 
\mbox{$\ket{\Phi(t)}\in{\mathfrak H}$}, we expand it into the instantaneous
system energy eigenbasis 
\mbox{${H}_{\rm{sys}}(t)\ket{\Psi_s(t)}=E_s(t)\ket{\Psi_s(t)}$} 
in ${\mathfrak H}_{\rm{sys}}$ via 
\mbox{$\ket{\Phi(t)}=
\sum_s a_s(t)\ket{\Psi_s(t)}\otimes\ket{\alpha_s(t)}$}, 
where $a_s$ are the corresponding amplitudes and 
$\ket{\alpha_s}\in{\mathfrak H}_{\rm{bath}}$ denote the associated  
(normalized but not necessary orthogonal) states of the reservoir.  
Insertion of this expansion into the Schr\"{o}dinger equation  
\mbox{$i\ket{\dot{\Phi}(t)}={H}(t)\ket{\Phi (t)}$} yields 
($\hbar=1$)
\bea   
\label{Diff}  
\pdiff{}{t} 
\left(a_s e^{i\varphi_s}\right) 
&=&  
e^{i\varphi_s}
\sum_{r\ne s} a_r
\left(   
\frac{\bra{\Psi_s }\dot{H}_{\rm{sys}}\ket{\Psi_r}}{\Delta E_{sr}} 
\,\bracket{\alpha_s}{\alpha_r}   
\right.
\nn
&&  
\left.
\phantom{\frac{ 1}{ 1}}
-i\lambda
\bra{\alpha_s}\bra{\Psi_s}{H}_{\rm{int}}\ket{\Psi_r}\ket{\alpha_r}
\right)
\,,            
\ea 	
with the energy gaps \mbox{$\Delta E_{sr}(t)=E_s(t)-E_r(t)$} of the
system and the total phase (including the Berry phase)
\bea
\label{Phase}
\varphi_s(t) 
&=& 
\int_0 ^t dt' 
\Big\{ 
E_s (t') +
H_{\rm{bath}}^{ss} (t') +
\lambda H_{\rm{int}}^{ss} (t') 
\nn
&&  
-i\bracket{\Psi_s(t')}{\dot{\Psi}_s(t')} 
-i\bracket{\alpha_s (t')}{\dot{\alpha}_s(t')}  
\Big\} 
\,, 
\ea     
containing the energy shift  
\mbox{$H_{\rm{int}}^{sr}=
\bra{\alpha_s}\bra{\Psi_s}{H}_{\rm{int}}\ket{\Psi_r}\ket{\alpha_r}$}
and
\mbox{$H_{\rm{bath}}^{sr}=
\bra{\alpha_s}{H}_{\rm{bath}}\ket{\alpha_r}$}.  
Evidently, there are two contributions for transitions in the Hilbert 
space ${\mathfrak H}_{\rm{sys}}$ of the system:
The first term on the right-hand side of Eq.~(\ref{Diff}) describes
the transitions caused by a non-adiabatic evolution 
(see, e.g., \cite{sarandy}). 
Note, however, that the factor $\bracket{\alpha_s}{\alpha_r}$ and the
additional phases in Eq.~(\ref{Phase}) give rise to extra terms in the
adiabatic expansion. 
The second term in Eq.~(\ref{Diff}) directly corresponds to transitions
caused by the interaction with the bath. 
Since we are mainly interested in the impact of the coupling to the
bath, we shall assume a perfectly adiabatic evolution 
\cite{channels}
of the system itself 
\mbox{$\bra{\Psi_s}\dot{H}_{\rm{sys}}\ket{\Psi_r}\ll\Delta E_{sr}^2$}
such that the first term in Eq.~(\ref{Diff}) is negligible and the
second one dominates. 


The quantum Ising chain of $n$ spins we are going to study exhibits a 
time-dependent nearest-neighbor interaction $g(t)$ plus transverse
field $B(t)=1-g(t)$ 
\bea
\label{Hamiltonian}
{H}_{\rm{sys}}(t) 
= 
-\sum_{j=1}^n\left\{[1-g(t)]\,\sigma^x_j+
g(t)\,\sigma^z_j\sigma^z_{j+1}\right\} 
\,, 
\ea
where ${\bm\sigma}_j=(\sigma^x_j,\sigma^y_j,\sigma^z_j)$ are the
spin-1/2 Pauli matrices acting on the $j$th qubit and periodic boundary
conditions \mbox{${\bm\sigma}_{n+1}={\bm\sigma}_1$} are imposed. 
Choosing $g(0)=0$ and $g(T)=1$ where $T$ is the evolution time,  
the system evolves from the paramagnetic state 
$\ket{\rightarrow\rightarrow\rightarrow\dots}$ 
trough a second-order \cite{sachdev} quantum phase transition at 
$g_{\rm cr}=1/2$ to 
the ferromagnetic phase $\ket{\uparrow\uparrow\uparrow\dots}$ or 
$\ket{\downarrow\downarrow\downarrow\dots}$. 


A major advantage of the above Hamiltonian is that it can be
diagonalized exactly \cite{sachdev}. 
Let us briefly review the main steps of the diagonalization of 
${H}_{\rm{sys}}$, where we switch temporarily to the Heisenberg
picture for convenience:
The set of $n$ qubits in (\ref{Hamiltonian}) can be mapped to a system
of $n$ spinless fermions $c_j$ via the Jordan-Wigner \cite{jordan} 
transformation given by  
\mbox{$\sigma^x_j=1-2c_j^{\dag}c_j$} 
and 
\mbox{$\sigma^z_j=-(c_j^{\dag}+c_j)\prod_{\ell < j}\sigma^x_\ell$}.  
In terms of the fermionic operators $c_j^{\dag}$ and $c_j$, the
Hamiltonian assumes a bilinear form containing 
\mbox{$(1-g)(1-2c_j^{\dag}c_j)$}  
and
\mbox{$g(c_{j+1}c_j + c_{j+1}^{\dag}c_j + 
c_j^{\dag}c_{j+1} + c_j^{\dag}c_{j+1}^{\dag})$}, i.e., the fermion
number $n_j=c_j^{\dag}c_j$ is not conserved in general. 
This bilinear form can now be diagonalized by a Fourier transformation  
\mbox{$c_j = \sum_k \tilde{c}_k \, e^{-i k (ja)} / \sqrt{n}$} followed
by a Bogoliubov \cite{bogoliubov} transformation 
\mbox{$\tilde{c}_k(t)=u_k(t)\,\gamma_k+iv_k^*(t)\,\gamma_{-k}^{\dag}$}. 
Since the new set of fermionic operators \mbox{$\gamma_k$} is supposed
to be time-independent, the Bogoliubov coefficients $u_k$ and
$v_k$ must satisfy \cite{dziarmaga} the equations of motion 
\mbox{$i\dot u_k = -\alpha_k u_k + \beta_k v_k$} 
and 
\mbox{$i\dot v_k = \alpha_k v_k + \beta_k u_k$},  
where 
\mbox{$\alpha_k(t) = 2 - 4 g(t) \cos^2(k a/2)$}
and 
\mbox{$\beta_k(t) =  2 g(t) \sin(k a)$}. 
For an adiabatic evolution 
\mbox{$\bra{\Psi_s}\dot{H}_{\rm{sys}}\ket{\Psi_r}\ll\Delta E_{sr}^2$},
these equations of motion can be solved approximately via 
\mbox{$u_k(t) \approx [\alpha_k(t)+\epsilon_k(t)]
\exp\{-i\int_0^t dt'\epsilon_k(t')\}/{\cal N}_k(t)$}
as well as 
\mbox{$v_k(t) \approx -\beta_k(t)
\exp\{-i\int_0^t dt'\epsilon_k(t')\}/{\cal N}_k(t) $}
with the normalization 
\mbox{${\cal N}_k=\sqrt{2\epsilon_k^2 + 2\alpha_k\epsilon_k}$} 
ensuring 
\mbox{$|u_k|^2 + |v_k|^2 = 1$} 
and the single-particle energies 
\bea
\label{single-particle}
\epsilon_k(t) = 2\sqrt{1-4g(t)\left[1-g(t)\right]\cos^2\left(ka/2\right)} 
\,.
\ea
All the excitation energies $\epsilon_k$ assume their minimum values
$\epsilon_k^{\rm min}=2|\sin(ka/2)|$ at the critical point 
$g_{\rm cr}=1/2$.   
In the following, we study the scaling of the involved quantities in
the continuum limit $n\uparrow\infty$. 
In view of the $k$-spectrum 
$k\in\pi(1+2{\mathbb Z})/(an)\;:\;|ka|<\pi$, where 
$a$ is the lattice spacing, the minimum gap scales as 
$\Delta E_{\rm min}=\ord(1/n)$. 
Finally, the Hamiltonian (\ref{Hamiltonian}) reads 
\bea
\label{HFinal}
{H}_{\rm{sys}}(t) = \sum_k \epsilon_k(t) 
\left(\gamma_k^{\dag}\gamma_k -\frac{1}{2} \right)
\,,
\ea
and hence its (instantaneous) ground state contains no fermionic
quasi-particles \mbox{$\forall_k{\gamma}_k\ket{\Psi_0(t)}=0$}. 
Without the environment, the number of fermionic quasi-particles 
$\gamma_k^{\dag}\gamma_k$ would be conserved and the system would stay
in an eigenstate (e.g., ground state) for an adiabatic evolution.


Of course, the impact of decoherence depends on the properties of the
bath and its interaction with the system (decoherence channels).
In order to derive generally applicable results, we do not specify the
bath ${H}_{\rm{bath}}$ in much detail and start with an interaction
$\lambda \, {H}_{\rm{int}}$ which is always present:  
In the Hamiltonian ${H}_{\rm{sys}}$ in Eq.~(\ref{Hamiltonian}), the
transverse field $B(t)=1-g(t)$ appears as a classical control
parameter $B_{\rm cl}$. 
However, the external field $B \to B_{\rm cl}+\delta B$ does also
possess (quantum) fluctuations $\delta B$, which couple to the system
of Ising spins.  
Therefore, we start with the following interaction Hamiltonian 
\bea
\label{Interaction1}
H_{\rm{int}} = \delta B\sum_j \sigma^x_j
\,,
\ea
where $\delta B$ denotes the reservoir operator. 
Incidentally, this interaction Hamiltonian yields the same matrix
elements as the non-adiabatic corrections 
$\bra{\Psi_s }\dot{H}_{\rm{sys}}\ket{\Psi_r}$
in Eq.~(\ref{Diff}), which can therefore be calculated analogously. 

Starting in the system's ground state $a_0(t=0)=1$, the excitations
$s>0$ caused by the weak interaction $\lambda H_{\rm{int}}$ with the
bath $\mathfrak{A}_s = a_s(T) \exp\{i\varphi_s(T)\}$ can be
calculated via response theory, i.e., the solution of Eq.~(\ref{Diff})
to first order in $\lambda\ll1$ is
\bea          
\label{Amplitude1}          
\mathfrak{A}_s 
\approx 
-i\lambda\int d\omega\,
f_s(\omega) 
\int_0^T dt\,
\bra{\psi_s(t)} \sum_j\sigma^x_j \ket{\psi_0(t)}
\times 
\nn
\times 
\exp\left\{i\left[-\omega t+\int_0^t dt'\,\Delta E_{s0}(t')\right]\right\}    
\,.
\ea       
We have subsumed all relevant properties of the environment into
the spectral function $f(\omega)$ of the bath 
\bea
\label{Fourier1}
e^{i\Delta\varphi_s(t)} 
\bra{\alpha_s(t)}\delta B(t)\ket{\alpha_0(t)}  
=           
\int d\omega\,e^{-i\omega t}f_s(\omega) 
\,,       
\ea
where $\Delta\varphi_s$ coincides with $\varphi_s-\varphi_0$ in 
Eq.~(\ref{Phase}) apart from the system's energy gap $\Delta E_{s0}$
and is typically dominated by the contribution from
\mbox{$H_{\rm{bath}}^{ss}-H_{\rm{bath}}^{00}$}.  
As a first approximation, we assume that $f(\omega)$ does not change
significantly if we increase the system size $n$ (scaling limit).

After inserting the Jordan-Wigner \cite{jordan} transformation, the
matrix element in Eq.~(\ref{Amplitude1}) reads 
\bea
\label{Element}
\sum_j\bra{\psi_s}\sigma^x_j(t)\ket{\psi_0} 
\approx
\frac{2ig(t)\sin(ka)}{\epsilon_k(t)}
\bra{\psi_s}{\gamma}_k^{\dag}{\gamma}_{-k}^{\dag}\ket{\psi_0} 
\,,
\ea
where the $\approx$ sign refers to the adiabatic approximation. 
Thus, it is only non-vanishing for excited states $\ket{\psi_s}$
containing two quasi-particles $s=(k,-k)$ with opposite momenta and
hence we get $\Delta E_{s0}=2\epsilon_k$. 
In order to solve the remaining time integrals, it is useful to
distinguish different $\omega$-regimes:
First of all, in order to have a {\em quantum} phase transition 
(or a working adiabatic quantum computer), the environment should be
cold enough to permit the preparation of the system in the initial
ground state, i.e., $\omega\ll2=\epsilon_k(t=0)$. 
For intermediate positive frequencies 
$2\gg\omega\gg\Delta E_{s0}^{\rm min}\approx2|ka|$, 
we may solve the time integral via the saddle-point 
(or stationary phase) approximation. 
The saddle-point condition for the exponent in Eq.~(\ref{Amplitude1})
reads $\omega=\Delta E_{s0}(t_*)=2\epsilon_k(t_*)$, which corresponds
to energy conservation. 
This condition yields two saddle points shortly before and after the
transition $g(t_*^\pm)\approx1/2\pm\sqrt{\omega^2-4k^2a^2}/8$. 
For the spectral excitation amplitude $\mathfrak{A}_s^\omega$ defined
via $\mathfrak{A}_s=\int d\omega\,f_s(\omega)\,\mathfrak{A}_s^\omega$,
the saddle-point approximation yields 
\bea
\label{spectral-Amplitude}    
\mathfrak{A}_s^{\omega\gg2|ka|}
=\ord\left(\frac{\lambda ka}{\sqrt{\omega\dot
    g(t_*)\sqrt{\omega^2-4k^2a^2}}}\right) 
\,.
\ea
Of course, the result depends on the interpolation dynamics $g(t)$,
see Table~\ref{Table}. 
For a constant speed interpolation $g(t)=t/T$, the run-time needed for
an adiabatic evolution scales as $T=\ord(n^2)$ due to the minimum gap 
$\Delta E_{\rm min}=\ord(1/n)$. 
For adapted interpolation dynamics $\dot g(t)\propto\Delta E(t)$
or $\dot g(t)\propto\Delta E^2(t)$, however, one may achieve shorter
run-times of $T=\ord(n\ln n)$ or $T=\ord(n)$, respectively
\cite{schaller2006b}.  
 
\begin{table}[h] 
\begin{tabular}{|c|c|c|}
\hline
&
$1\gg\omega\gg2ka$
&
$1\gg\omega\approx2ka$
\\
\hline
$\ddot g(t)=0$
&
$\ord(\lambda ka\omega^{-1}n)$ 
&
$\ord(\lambda n^2\omega\ln\omega)$ 
\\
\hline
$\dot g(t)\propto\Delta E(t)$
&
$\ord(\lambda ka\omega^{-3/2}\sqrt{n})$
&
$\ord(\lambda n\ln n)$
\\
\hline
$\dot g(t)\propto\Delta E^2(t)$
&
$\ord(\lambda ka\omega^{-2})$
&
$\ord(\lambda n)$ 
\\
\hline
\end{tabular}
\caption{\label{Table} Scaling of the spectral excitation amplitude
  $\mathfrak{A}_s^\omega$ in the saddle-point approximation
  ($\omega\gg2ka$) and its upper bound ($\omega\approx2ka$) for
  different interpolation dynamics $g(t)$, where $\Delta
  E(t)=2\epsilon_{k=\pi/(an)}(t)$ denotes the fundamental gap. In all
  cases, the total excitation probability (sum over all $\omega$ and
  $k$) increases with system size $n$.}
\end{table}

\begin{figure}
\includegraphics[height=5cm]{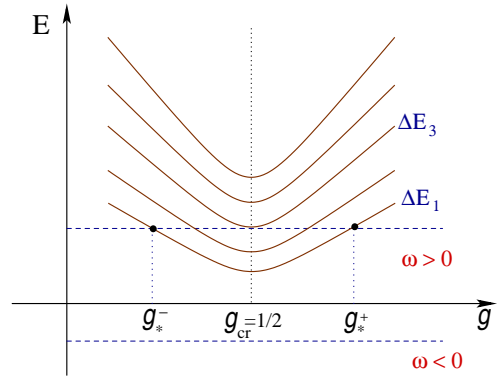} 
\caption{\label{Energy} [Color online] Sketch of the excitation
  spectrum of the Ising  
  chain ${H}_{\rm{sys}}$ as a function of $g$. For a given frequency
  $\omega>0$, real saddle points correspond to intersections of the
  (solid) energy level curves (e.g., $\Delta E_1$) with the (dashed)
  vertical $\omega$-line which occur shortly before ($g_*^-$) and
  after ($g_*^+$) the quantum phase transition at $g_{\rm
  cr}=1/2$. The saddle-point approximation can 
  only be applied if the intersection angle is large enough, i.e., for
  the drawn $\omega>0$ line, it would work for $\Delta E_1$, but not
  for $\Delta E_3$ etc.   
} 
\end{figure} 

The next (higher-order) terms of the saddle-point expansion scale
with $\ord(\lambda \dot g(t_*)[\omega^2-4k^2a^2]^{-1/2}/\omega)$ and hence the
saddle-point approximation breaks down if $\omega$ approaches the
minimum gap $\Delta E_{s0}^{\rm min}\approx2|ka|$, see
Fig.~\ref{Energy}. 
In this case, we may obtain an upper bound for the time integral in 
Eq.~(\ref{Amplitude1}) via omitting all phases, see Table~\ref{Table}.  
For frequencies far below $2|ka|$, the saddle points at  
$g(t_*^\pm)\approx1/2\pm\sqrt{\omega^2-4k^2a^2}/8$ move away from the
real axis and thus the amplitudes are exponentially suppressed in the
adiabatic limit
\bea
\label{suppressed} 
\mathfrak{A}_s^{\omega\ll2|ka|}
=\ord\left(\lambda\exp\left\{-T(ka)^2/2\right\}\right)
\,,
\ea
for $g(t)=t/T$ and similarly for the other interpolations. 
Finally, for negative frequencies $\omega<0$, the saddle points
collide with the branch cut generated by the square-root in  
$\epsilon_k$.
By deforming the integration contour into the complex plane, it can be
shown via an argumentation analogous to \cite{schaller2006b} that the
amplitudes are also exponentially suppressed in this case. 
This result can be understood in the following way:
For frequencies $\omega$ below the lowest excitation energies, the
energy $\omega$ of the reservoir modes is not sufficient for exciting 
the system via energy-conserving transitions. 
Hence excitations can only occur via non-adiabatic processes 
for which energy-conservation becomes ill-defined, but these
processes are suppressed if the evolution is slow enough.  


In summary, we studied the quantum phase transition from paramagnetic 
to ferromagnetic phase in the quantum Ising chain in a transverse
field via its analytical diagonalization and calculated the excitation
probabilities \cite{channels} 
caused by a weak coupling to a rather general environment 
(including possible non-perturbative behavior of the reservoir). 
Since the Ising model is considered \cite{sachdev} a prototypical
example for a second-order quantum phase transition, we expect our
results to reflect general features of second-order transitions.    
For the decoherence channel (\ref{Interaction1}) which is always
present (though possibly not the dominant channel), we already found
that the total excitation probability increases with system size $n$
(continuum limit): 
Even though the probability for the {\em lowest} excitation
$k=\pm\pi/(an)$ can be kept under control for a bath which is
well-behaved in the infra-red limit (see also \cite{markus}), the
existence of {\em many} excited states 
$k\in\pi(1+2{\mathbb Z})/(an)\;:\;|ka|<\pi$
converging near the critical point causes the growth of the error
probability for large systems.  
This growth can be slowed down a bit via adapted interpolation schemes
$g(t)$, but not stopped. 
Other decoherence channels will display the same general behavior:
E.g., for $\omega\gg|ka|$, the associated amplitudes scale as 
$\mathfrak{A}_s^\omega=\ord(\lambda\phi_s(t_*)/\sqrt{\dot g(t_*)})$,
where $\phi_s$ denotes the matrix element in analogy to
(\ref{Element}). 
Typically, for a homogeneous coupling to the bath, $\phi_s$ does not
strongly depend on the system size $n$ (for given $ka$ and $\omega$). 
Since $\dot g(t_*)$ decreases for $n\uparrow\infty$ or at least
remains constant [for $\dot g(t)\propto\Delta E^2(t)$], the total
excitation probability again increases with system size $n$ 
\cite{few}. 

Using the analogy between adiabatic quantum algorithms and quantum
phase transitions \cite{latorre,gernot}, this result suggests
scalability problems of the corresponding adiabatic quantum algorithm
-- unless the temperature of the bath stays below the ($n$-dependent)
minimum gap \cite{childs_robust} or the coupling to the bath decreases
with increasing $n$. 
These problems are caused by the accumulation of many levels at the
critical point $g=1/2$, 
which presents the main difference to isolated avoided level crossings
(corresponding to first-order phase transitions) discussed earlier
\cite{markus}. 
It also causes some difficulties for the idea of thermally assisted
quantum computation (see, e.g., \cite{amin}) since, in the presence of
too many available levels, the probability of hitting the ground
state becomes small.

Therefore, in order to construct a scalable adiabatic quantum
algorithm in analogy to the Ising model, suitable error-correction
methods will be required. 
As one possibility, one might exploit the quantum Zeno effect 
and suppress transitions in the system by constantly measuring the
energy, see for example \cite{childs_zeno}. 
As another interesting idea, let us study a spatial sweep through the 
phase transition, i.e., we do not cross the critical point in a
homogeneous way, but adopt the following step-wise interpolation:   
Starting from the initial Hamiltonian 
$\sigma^x_1+\sigma^x_2+\sigma^x_3+\sigma^x_4+\dots$, we change it
slowly to 
$\sigma^z_1\sigma^z_2+\sigma^x_3+\sigma^x_4+\dots$
and afterwards to 
$\sigma^z_1\sigma^z_2+\sigma^z_2\sigma^z_3+\sigma^x_4+\dots$
etc. 
This corresponds to a nonlinear interpolation path between the two
Hamiltonians. 
In this case, the minimum gap 
(in the relevant subspace that is even under bit flip) 
remains independent of the system size
$n$ and the run-time $T$ scales linear in $n$ (number of steps). 
Hence, decoherence could be strongly suppressed for a
low-temperature bath.  
Of course, the generalization of all these concepts and results to
more interesting cases such as the (NP-complete) two-dimensional Ising
model is highly non-trivial and requires further investigations.

This work was supported by the Emmy-Noether Programme of the
German Research Foundation (DFG) under grant SCHU~1557/1-2 
and by grant SCHU~1557/2-1.

$^*$ email: {\tt schuetz@theory.phy.tu-dresden.de}



\begin{thebibliography}{9999}
\bibitem{channels}
Without the coupling to the environment $\lambda=0$, the system would
stay in its ground state. 
%
Thus the only decoherence channel available is heating 
(i.e., excitations), the phase damping and decay channels, for
example, play no major role here.
\bibitem{few}
If only a few spins are coupled to the environment, the
matrix element $\phi_s$ (for given $ka$ and $\omega$) will generically 
decrease $\phi_s=\ord(1/n)$ and then the error probability may be kept
under control [for $\dot g(t)\propto\Delta E^2(t)$].
\end{thebibliography}
\end{document}